\documentclass[12pt]{iopart}
\usepackage{iopams}
\usepackage{graphicx}
\newcommand{\veps}{\varepsilon}
\newcommand{\rhm}{\rho_{\mathrm{m}}}
\newcommand{\prm}{p_{\mathrm{m}}}
\newcommand{\dph}{\phi'}
\newcommand{\pat}{\partial}
\newcommand{\rhmz}{\rho_{\mathrm{m}0}}
\newcommand{\prmz}{p_{\mathrm{m}0}}
\begin{document}
\title[Analysis of regular inflationary cosmological models in Poincar\'e gauge theory of gravity]%
{Analysis of regular inflationary cosmological models with two torsion functions in Poincar\'e
gauge theory of gravity}


\author{A.S. Garkun$^1$, V.I. Kudin$^2$ and A.V. Minkevich$^{1,3}$}

\address{$^1$Department of Theoretical Physics, Belarussian State University, Belarus}
\address{$^2$Department of Technical Physics, Belarussian National Technic University, Belarus}
\address{$^3$Department of Physics and Computer Methods, Warmia and Mazury University in Olsztyn,
Poland}
 \eads{\mailto{minkav@bsu.by}, \mailto{garkun@bsu.by}}

\begin{abstract}
Analysis of regular inflationary cosmological models with two
torsion functions filled with scalar field with quadratic
potential and ultrarelativistic matter is carried out numerically.
Properties of  different stages of regular inflationary
cosmological solutions are studied, restrictions on admissible
values of parameters and initial conditions at transition from
compression to expansion are found. The structure of extremum
surface in space of physical variables is investigated.
\end{abstract}
 \submitto{\CQG}

\section{Introduction}

As it was shown in a number of papers (see [1-5] and references
herein), the Poincar\'e gauge theory of gravity (PGTG) offers
opportunities to solve some principal problems of general
relativity theory (GR) and modern cosmology. It is because the
gravitational interaction in the framework of PGTG can have the
repulsive character under certain physical conditions in usual
gravitating systems with positive values of energy density and
pressure. As a result the PGTG allows to build totally regular Big
Bang scenario with an accelerating stage of cosmological expansion
at present epoch. Such scenario is based on homogeneous isotropic
models (HIM) with two torsion functions built in the framework of
PGTG in our previous paper [4], where as an example a particular
regular inflationary cosmological solution was obtained by
numerical calculations.

The present paper is devoted to analysis of regular inflationary
cosmological HIM with two torsion functions in the PGTG. In
Section 2 cosmological equations for such models in dimensionless
form are given. In Section 3 the most important general properties
of regular inflationary cosmological solutions are analyzed
numerically. In Section 4 we study extremum surfaces, on which the
Hubble parameter vanishes and initial conditions for investigated
solutions are given.

\section{Cosmological equations for HIM with two torsion functions in dimensionless form}

In the framework of PGTG any HIM is described by three geometric
characteristics: the scale factor of Robertson-Walker metrics $R$
and two torsion functions $S_{1}$ and $S_{2}$ as functions of time
$t$. By using general form of gravitational Lagrangian including
both a scalar curvature and various invariants quadratic in
gravitational field strengths --- curvature and torsion tensors,
cosmological equations for HIM with two torsion functions were
deduced in [4]. The energy density $\rho$ and pressure $p$ of
gravitating matter play the role of sources of gravitational field
in cosmological equations. These equations contain three
indefinite parameters: parameter $\alpha$ with inverse dimension
of energy density determining the scale of extremely high energy
densities, parameter $b$ with dimension of parameter $f_0=(16\pi
G)^{-1}$  ($G$ is Newton's gravitational constant, the light
velocity $c=1$) and dimensionless parameter $\veps$. As was shown
in [4], by certain restrictions on indefinite parameters
cosmological equations for HIM lead to observable accelerating
cosmological expansion at present epoch. According to obtained
estimations the parameter $b$ has to be close to $f_0$ and  the
parameter $\veps$ satisfies the following conditions $0<\veps\ll
1$.

In order to analyze inflationary cosmological solutions we will
consider HIM filled with scalar field $\phi$ with a potential $V$
and usual gravitating matter (values of gravitating matter are
denoted by means of index ``${m}$''). In this case we have:
$$
\rho=\frac{1}{2}\left(\frac{\pat\phi}{\pat t}\right)^2+V+\rho_m
    \quad (\rho>0), \quad
p=\frac{1}{2}\left(\frac{\pat\phi}{\pat t}\right)^2-V+p_m.
$$
To investigate inflationary solutions we transform cosmological
equations obtained in [4] to dimensionless form by introducing
dimensionless units for all variables and parameter $b$ entering
these equations and denoted by means of \ $\tilde{}$\, by the
following way:
\begin{equation}\label{dimless}
\begin{array}{lcl}
    t\to\tilde{t}=t/\sqrt{6 f_0 \alpha},& \qquad\qquad
            & R\to\tilde{R}=R/\sqrt{6f_0 \alpha},\\
    \rho\to\tilde{\rho}=\alpha\,\rho, & & p\to\tilde{p}=\alpha\,p,\\
    \phi\to \tilde{\phi} = \phi/\sqrt{6f_0}, & &
            b\to\tilde{b} = b/f_0, \\
    H\to\tilde{H}=H\sqrt{6f_0 \alpha}, & &
            S_{1,2}\to\tilde{S}_{1,2}=S_{1,2}\sqrt{6f_0 \alpha}.
\end{array}
\end{equation}
where dimensionless Hubble parameter $\tilde{H}$ is defined  by
usual way $\tilde{H}=\tilde{R}^{-1}\frac{d \tilde{R}}{d
\tilde{t}}$. Numerical analysis in Section~3 will be made by
choosing quadratic potential $V=\frac{1}{2}m^2\phi^2$, then in
accordance with (1) the transition to dimensionless units gives:
$$
V\to \tilde{V}=\alpha V, \qquad
\tilde{V}=\frac{1}{2}\tilde{m}^2\tilde{\phi}^2, \qquad
m\to\tilde{m}=m\sqrt{6f_0\alpha}.
$$
Because of the transformation (1) cosmological equations
((22)--(23) in [4]) take the following dimensionless form, where
the differentiation with respect to dimensionless time $\tilde{t}$
is denoted by means of the prime and the sign of \ $\tilde{}$\, is
omitted below:
\begin{eqnarray}\label{22}\fl
    \frac{k}{R^2} + (H-2S_1)^2= \frac{1} {Z}
        \left[
            {\rho  +\left(Z- b\right) S_2^2
            + \frac{1}{4} \left( {\rho  - 3p - 12bS_2^2 } \right)^2 }
        \right]
\nonumber\\
        - \frac{\varepsilon}{2Z}
            \left[
                {\left( {HS_2  + S_2' } \right)^2
                + 4\left( {\frac{k}{{R^2 }} - S_2^2 } \right)S_2^2 }
            \right],
\end{eqnarray}
\begin{eqnarray}\label{23}\fl
    H'+H^2-2HS_1-2S_1' = -\frac{1} {2Z}
        \left[
            \rho  + 3p - \frac{1 } {2} \left( {\rho  - 3p - 2bS_2^2 } \right)^2
        \right]
\nonumber\\
        - \frac{\varepsilon }{Z}\left( {\rho  - 3p - 2bS_2^2 } \right)S_2^2
        + \frac{{\varepsilon }} {2Z}
            \left[ {\left( {HS_2  + S_2' } \right)^2
                + 4\left( {\frac{k}{{R^2 }} - S_2^2 } \right)S_2^2 }
            \right],
\end{eqnarray}
($Z\equiv 1+\rho - 3p - 2\left( {b + \varepsilon }
\right)S_2^2 = 1 + 4V-\dph^2+\rhm-3\prm- 2\left( {b + \varepsilon }
\right)S_2^2$).
 The torsion function $S_{1}$ in dimensionless form entering (2)--(3) is
\begin{equation} \label{20}
    S_1  = -\frac{3}{4Z} \left[
            2\frac{\pat V}{\pat \phi} \phi' \ +H\left({Y + 2\dph^2}\right)
            -\frac{2}{3}\left( {2b - \veps } \right) S_2 \, S_2'
        \right],
\end{equation}
where
\[
    Y \equiv \left(\rhm+\prm\right)%
        \left(3\frac{d\prm}{d\rhm}-1 \right) + 2\varepsilon S_2^2
\]
and  dimensionless torsion function $S_{2}$ satisfies the following differential equation of the
second order:
\begin{eqnarray}\label{21}\fl
    \varepsilon \left[ S_2''  + 3H S_2'  + 3H' S_2  - 4\left(S_1'  - 3 HS_1
        + 4S_1^2\right) S_2  \right]
\nonumber \\
        - 2\left( {\rho  - 3p - 2bS_2^2 } \right)S_2
        - 2\left( {1  - b}\right)S_2  = 0\,.
\end{eqnarray}
By using dimensionless units the equation for scalar field and conservation law for gravitating
matter have the usual form
\begin{eqnarray}
\label{14}
& & \phi''+3H\dph=-\frac{\pat V}{\pat\phi}\, , \\
\label{15}
& & \rho_m'+3H\left(\rhm+\prm\right)=0.
\end{eqnarray}
By means of (6)--(7) and (4) for $S_{1}$-function we can transform cosmological equations (2)--(3)
and (5) for $S_{2}$-function to the following form:
\begin{eqnarray}\label{27}\fl
    H^2\left\{
            \left[ Z+\frac{3}{2}\left( Y+2\dph \right) \right]^2
            +\frac{1}{2}\veps S_2^2 Z
        \right\}
\nonumber\\
    +6 H\left\{
            \left[ Z+\frac{3}{2}\left( Y+2\dph^2 \right) \right]
            \times\left[
                \frac{\pat V}{\pat \phi}\dph
                -\frac{1}{3}\left( 2b-\veps \right) S_2 S_2'
            \right]
            +\veps S_2 S_2' Z
        \right\}
\nonumber\\
    +9\left[
            \frac{\pat V}{\pat \phi}\dph-\frac{1}{3}\left( 2b-\veps \right)S_2 S_2'
        \right]^2
    +\frac{1}{2}\veps \left[
            S_2'^2+4\left( \frac{k}{R^2}-S_2^2 \right) S_2^2
        \right] Z
\nonumber\\
    - \left[
            \rhm+\frac{1}{2}\dph+V-b S_2^2+\frac{1}{4}
                \left(
                    \rhm-3\prm+4V-\dph^2-2b S_2^2
                \right)^2
        \right] Z
\nonumber\\
    +\left( \frac{k}{R^2}-S_2^2 \right) Z^2=0,
\end{eqnarray}
\begin{eqnarray}\label{28}\fl
    H'\left[
            1 + \frac{3}{2Z}\left( Y + 2\dph^2 \right)
        \right]
    +H^2\left\{
            1 + \frac{3}{2Z}\left( Y + 2\dph^2 \right)
            -\frac{9}{2Z^2}\left( Y + 2\dph^2 \right)
                \left( Y + 2\dph^2 - 2 \veps S_2^2 \right)
\right.\nonumber\\
\left.
            -\frac{9}{2 Z}\left[
                    3\frac{d^2 \prm}{d\rhm^2} \left( \rhm + \prm \right)^2
                    +\left( 3 \frac{d\prm}{d\rhm}-1 \right)
                        \left( 1 + \frac{d\prm}{d\rhm} \right) \left( \rhm + \prm \right)
                    +4\dph^2
                \right]
        \right\}
\nonumber\\
    -\frac{3}{Z} H\left\{
            \left[4 \frac{\pat V}{\pat \phi}\dph +  \frac{1}{3}\left( 2b - 7\veps \right)S_2S_2' \right]
            +\frac{3}{Z}\left[
                    \left(
                        \frac{\pat V}{\pat\phi}\dph - \frac{1}{3}\left( 2b - \veps \right)S_2S_2'
                    \right)
\right.\right.\nonumber\\
\left.\left.\times
                    \left(Y+2\dph^2 - 2\veps S_2^2\right)
                    +\left( Y + 2\dph^2 \right)
                            \times\left(
                                \frac{\pat V}{\pat\phi}\dph
                                -\frac{2}{3}\left( b + \veps \right)S_2S_2'
                            \right)
                \right]\right\}
\nonumber\\
    +\frac{3}{Z}\left\{
            \frac{\pat^2 V}{\pat\phi^2}\dph^2-\left( \frac{\pat V}{\pat\phi} \right)^2
\right.\nonumber\\
\left.
            -\frac{6}{Z}\left(
                    \frac{\pat V}{\pat\phi}\dph - \frac{1}{3}\left( 2b-\veps \right) S_2S_2'
                \right)
                \times\left(
                    \frac{\pat V}{\pat\phi}\dph - \frac{2}{3}\left( b+\veps \right) S_2S_2'
                \right)
\right.\nonumber\\
\left.
            -\frac{1}{3}\left( 2b - \veps \right) \left(S_2'^2 + S_2 S_2'' \right)
        \right\}
\nonumber\\
    =-\frac{1}{2 Z}\left[
            \rhm + 3\prm -2\left( V - \dph^2 \right)
            - \frac{1}{2} \left( \rhm - 3\prm + 4V - \dph^2 - 2b S_2^2 \right)^2
        \right]
\nonumber\\
    -\frac{\veps}{Z}\left( \rhm - 3\prm + 4V - \dph^2 - 2b S_2^2 \right) S_2^2
\nonumber\\
    +\frac{\veps}{2Z}\left[
            \left( H S_2^2 + S'^2 \right)^2 + 4\left( \frac{k}{R^2} - S_2^2 \right) S_2^2
        \right],
\end{eqnarray}
\begin{eqnarray}\label{29}\fl
    S_2''\left[1 - \frac{2}{Z}\left( 2b - \veps \right) S_2^2 \right]
    +3H' S_2\left[1 + \frac{1}{Z}\left( Y + 2\dph^2 \right) \right]+3H S_2'
\nonumber\\
    -\frac{9}{Z} H^2S_2\left[
            Y + 6\dph^2 + 3\frac{d^2\prm}{d\rhm^2}\left( \rhm + \prm \right)^2
\right.\nonumber\\
\left.
            +\left( 3\frac{d\prm}{d\rhm} - 1 \right)\left( 1 + \frac{d\prm}{d\rhm} \right)
                \left( \rhm + \prm \right)
            +\frac{1}{Z}\left( Y + 2\dph^2 \right)
\right.\nonumber\\
\left.\vphantom{\frac{d\prm}{d\rhm}}
\times
                \left( Y + 2\dph^2 - 2\veps S_2^2\right)
        \right]
    -3H S_2\left\{
         \frac{4}{Z} \left(
                4\frac{\pat V}{\pat\phi}\dph - \frac{1}{2}\left( 2b + \veps \right) S_2S_2'
            \right)
\right.\nonumber\\
\left.
    +\frac{6}{Z^2} \left[
            \left(
                \frac{\pat V}{\pat\phi}\dph - \frac{1}{3}\left( 2b - \veps \right) S_2S_2'
            \right)
            \left( Y + 2\dph^2 - 2\veps S_2 S_2'\right)
\right.\right.\nonumber\\
\left.\left.
            -\left( Y + 2\dph^2 \right)\left(
                    \frac{\pat V}{\pat\phi}\dph - \frac{2}{3}\left( b + \veps \right) S_2S_2'
                \right)
            \right]\right\}
\nonumber\\
    -\frac{9}{Z^2} S_2\left[
            H\left(Y + 2\dph^2 \right)
            +2\left(
                    \frac{\pat V}{\pat\phi}\dph - \frac{1}{3}\left( 2b - \veps \right) S_2S_2'
                \right)
        \right]^2
\nonumber\\
    -\frac{6}{Z} S_2\left[
            \left( \frac{\pat V}{\pat\phi} \right)^2 - \frac{\pat^2 V}{\pat\phi^2}\dph^2
            +\frac{1}{3}\left( 2b - \veps f_0 \right) S_2'^2
\right.\nonumber\\
\left.
            +\frac{6}{Z}\left(
                    \frac{\pat V}{\pat\phi}\dph - \frac{1}{3}\left( 2b - \veps \right) S_2S_2'
                \right)
                \left(
                    \frac{\pat V}{\pat\phi}\dph - \frac{2}{3}\left( b + \veps \right) S_2S_2'
                \right)
        \right]
\nonumber\\
    -\frac{1}{\veps}\left[
            2\left( \rhm - 3\prm + 4V - \dph^2 - 2b S_2^2 \right)
            +2\left(1 - b\right)
        \right] S_2
    =0.
\end{eqnarray}
The cosmological equation (8) leads to the following equation for
extremum surface in space of variables ($\phi$, $\phi'$, $S_2$,
$S_2'$, $\rhm$), in points of which $H=0$:
\begin{eqnarray}\label{32}\fl
    \left[
            \rhmz + \frac{1}{2}{\dph_0}^2 + V_0 - b S_{20}^2
            +\frac{1}{4}\left(
                    \rhmz - 3\prmz + 4V_0 - {\dph_0}^2 - 2b S_{20}^2
                \right)^2
        \right] Z_0
\nonumber\\
    -\left[
            3\left( \frac{\pat V}{\pat\phi} \right)_{\!0}\dph_0
            - \left( 2b - \veps \right) S_{20}S_{20}'
        \right]^2
\nonumber\\
    -\frac{1}{2}\veps \left[
            {S_{20}'}^2 + 4\left( \frac{k}{R_0^2} - S_{20}^2 \right) S_{20}^2
        \right] Z_0
    -\left( \frac{k}{R_0^2} - S_{20}^2 \right) Z_0^2
    =0,
\end{eqnarray}
where variables on extremum surface are denoted by means of index
$0$. Then from (9) we obtain the following expression for
derivative of the Hubble parameter $H_0'$ in points of extremum
surfaces
\begin{eqnarray}\label{33}\fl
    H_0' Z_0^2\left\{
            1 + \left[
                    \rhmz - 3\prmz
                    +\frac{3}{2}\left( 3\left( \frac{d\prm}{d\rhm} \right)_{\!0} - 1 \right)
                        \left( \rhmz + \prmz \right)
                    +4V_0 + 2{\dph_0}^2
                \right]
        \right\}
\nonumber\\
    =\left[
            1+\left( \rhmz - 3\prmz + 4V_0 - {\dph_0}^2 - 6b S_{20}^2 \right)
        \right]
    \times\left\{
            3Z_0\left[
                    \frac{1}{2}\left( \rhmz - \prmz \right) + V_0
\right.\right.\nonumber\\
\left.\left.
                    - \frac{2}{3}bS_{20}^2
                    + \frac{1}{4} \left(
                            \rhmz - 3\prmz + 4V_0 - {\dph_0}^2 - 2b S_{20}^2
                        \right)^2
                \right]
\right.\nonumber\\
\left.
            +3 Z_0\left[
                    \left( \frac{\pat V}{\pat\phi} \right)_{\!0}^2
                    -\left( \frac{\pat^2 V}{\pat\phi^2} \right)_{\!0}{\dph_0}^2
                    +\frac{1}{6}\left( 4b - 3\veps \right) {S_{20}'}^2
                    -\frac{2}{3}\veps \left( \frac{k}{R_0^2} - S_{20}^2
                        \right) S_{20}^2
\right.\right.\nonumber\\
\left.\left.
                    -\frac{1}{3}\veps\left(
                            \rhmz - 3\prmz + 4V_0 - {\dph_0}^2 - 2b S_{20}^2
                        \right) S_{20}^2
                \right]
            -2\left( \frac{k}{R_0^2} - S_{20}^2 \right) Z_0^2
\right.\nonumber\\
\left.
            -18\veps S_{20} {S_{20}'}\left[
                    \left( \frac{\pat V}{\pat\phi} \right)_{\!0}\dph_0
                    -\frac{1}{3}\left( 2b - \veps \right) S_{20}{S_{20}'}
                \right]
        \right\}
\nonumber\\
    +\left( 2b - \veps \right) S_{20}^2\left\{
            72\left[
                    \left( \frac{\pat V}{\pat\phi} \right)_{\!0}\dph_0
                    -\frac{1}{3}\left( 2b - \veps \right) S_{20}{S_{20}'}
                \right]
\right.\nonumber\\
\left.
            \times\left[
                    \left( \frac{\pat V}{\pat\phi} \right)_{\!0}\dph_0
                    -\frac{1}{6}\left( 4b + \veps \right) S_{20}{S_{20}'}
                \right]
            +6 Z_0\left[
                    \left( \frac{\pat V}{\pat\phi} \right)_{\!0}^2
                    -\left( \frac{\pat^2 V}{\pat\phi^2} \right)_{\!0}{\dph_0}^2
\right.\right.\nonumber\\
\left.\left.
                    +\frac{1}{3}\left( 2b - \veps \right) {S_{20}'}^2
                \right]
\right.\nonumber\\
\left.
            +\frac{1}{\veps} Z_0^2\left[
                    2\left(
                            \rhmz - 3\prmz + 4V_0 - {\dph_0}^2 - 2b S_{20}^2
                        \right)
                    +2\left(1 - b\right)
                \right]
        \right\}.
\end{eqnarray}
The transition from compression stage to expansion stage takes
place on extremum surface. In the case of HIM filled at the
beginning of cosmological expansion with scalar field with
quadratic potential $V$ and ultrarelativistic matter
($\prm=\frac{1}{3}\rhm$) analyzed below the equation (11) for
extremum surface takes the following form:
\begin{eqnarray}\label{32a}\fl
    \left[
            \rhmz + \frac{1}{2}{\dph_0}^2 + \frac{1}{2}m^2\phi_0^2 - b S_{20}^2
            +\left(
                    2V_0 - \frac{1}{2}{\dph_0}^2 - b S_{20}^2
                \right)^2
        \right] Z_0
    -\left[
            3m^2\phi_0\dph_0
            - \left( 2b - \veps \right) S_{20}S_{20}'
        \right]^2
\nonumber\\
    -\frac{1}{2}\veps \left[
            {S_{20}'}^2 + 4\left( \frac{k}{R_0^2} - S_{20}^2 \right) S_{20}^2
        \right] Z_0
    -\left( \frac{k}{R_0^2} - S_{20}^2 \right) Z_0^2
    =0.
\end{eqnarray}

\section{Numerical analysis of regular inflationary cosmological solutions}

We will obtain cosmological solutions by integrating the system of
differential equations (9), (10), (6), (7) and by choosing initial
conditions for independent physical variables given on extremum
surface in accordance with (13). Because the most important
properties of cosmological inflationary solutions at the beginning
of cosmological expansion are connected with the presence of
scalar fields, at first we will analyze HIM filled by scalar field
without other gravitating matter ($\rhm=0$). For simplicity we
will consider flat HIM ($k=0$). By taking into account
restrictions on indefinite parameters leading to cosmological
acceleration at asymptotics [4], we will use $b=1$ for numerical
calculations.

Any regular inflationary cosmological solution includes the
following stages: the compression stage, the transition from
compression to expansion, the inflationary and post-inflationary
stages. Properties of various stages of regular inflationary
cosmological solution, generally speaking, depend on parameters
$\veps$ and $m$ and initial conditions for independent physical
variables. Because of relation (13) only three variables from the
following four quantities ($\phi_0$, $\phi'_0$, $S_{20}$,
$S_{20}'$) are independent. We will give the initial conditions at
the moment $t=0$ for ($\phi_0$, $S_{20}$, $S_{20}'$), then the
initial value of derivative $\phi_0'$ will be determined from
(13). At first, we will study the transition stage from
compression to expansion. The temporal behaviour of the Hubble
parameter $H(t)$, scalar field $\phi(t)$, torsion function
$S_2(t)$ and its derivative $S_2'(t)$ at this stage are presented
in Figures~1--3. The graphs in Figures~1--3 are obtained at the
following values of parameters $\veps=10^{-5}$, $m=1{.}1$ and
initial conditions: $\phi_0=10$, $S_{20}=0$ and $S_{20}'=10^{-4}$.
The characteristic feature of transition stage is essentially
non-linear oscillating behaviour of  the Hubble parameter.  There
is the correlation between the frequency of such oscillations and
that of the function $S_2'$ (see Fig.~1 and Fig.~4).

\begin{figure}[hbt!]
\begin{minipage}{0.48\textwidth}\centering{
 \includegraphics[width=\linewidth]{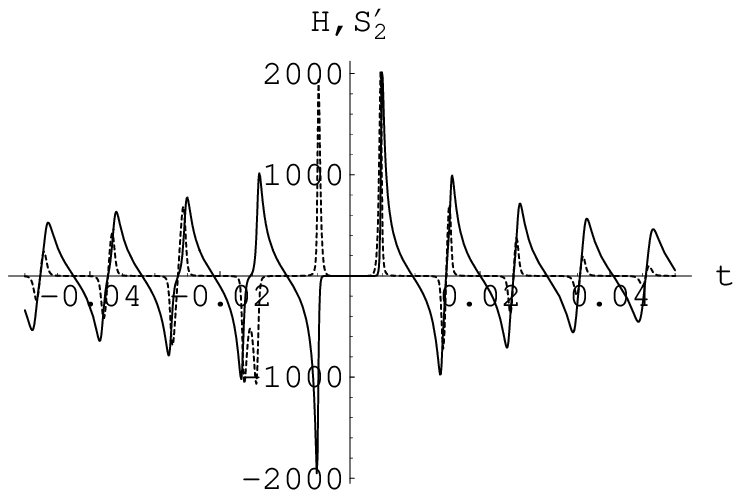}
} \caption[]{The temporal behaviour of $H(t)$ (firm line) and
$S_2'(t)$ (dotted line) at the transition stage ($S_2'$ is
normalized by the factor 0{.}1).}
\end{minipage}\, \hfill\,
\begin{minipage}{0.48\textwidth}\centering{
 \includegraphics[width=\linewidth]{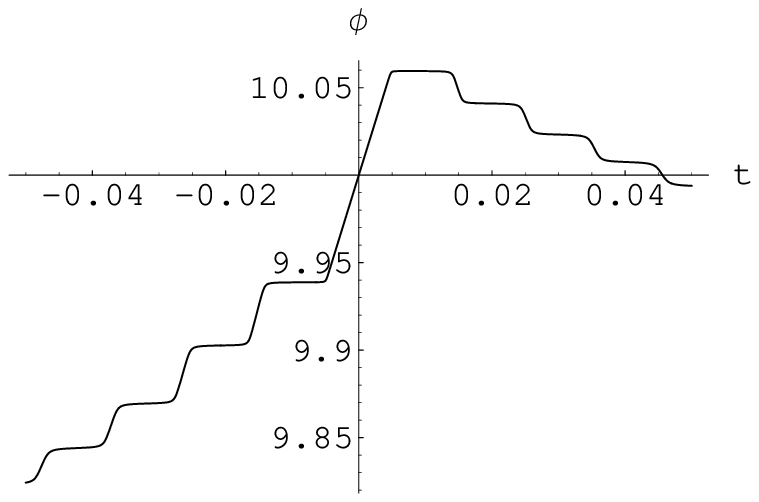}
} \caption[]{The temporal behaviour of $\phi(t)$ at the transition
stage.}
\end{minipage}\\
\end{figure}

\begin{figure}[hbt!]
\begin{minipage}{0.48\textwidth}\centering{
 \includegraphics[width=\linewidth]{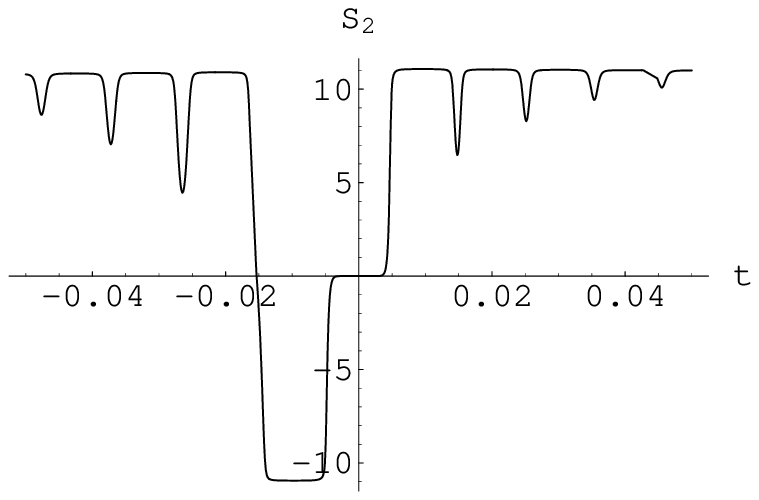}
} \caption[]{The temporal behaviour of $S_2(t)$ at the transition
stage.}
\end{minipage}\, \hfill\,
\begin{minipage}{0.48\textwidth}\centering{
 \includegraphics[width=\linewidth]{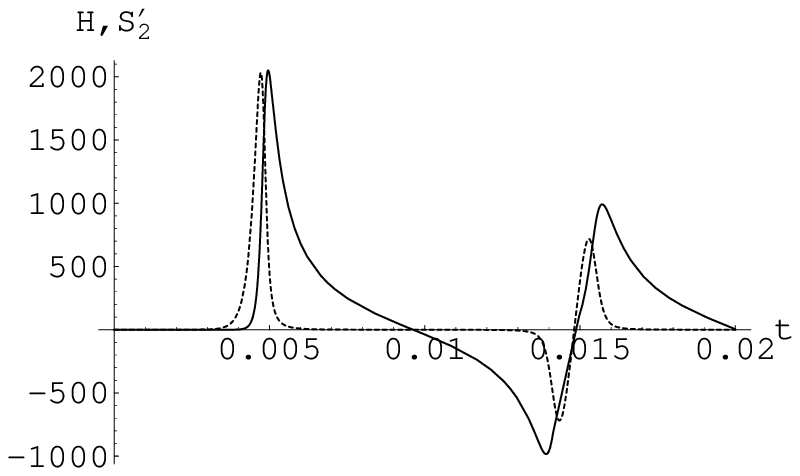}
} \caption[]{The correlation between $H$- and $S_2'$-oscillations
in the middle of the transition stage.}
\end{minipage}\\
\end{figure}

The oscillating behaviour of the Hubble parameter at transition
stage allows to estimate its duration only approximately. As
numerical analysis shows, the duration of transition stage is
smaller in one order or more than duration of inflationary stage.
At the end of transition stage the amplitude of $H$-oscillations
decreases with time, the value of $H$ becomes positive, and at
some moment the transition to inflationary stage with slow rolling
regime of scalar field  takes place. The inflationary stage is
finished at $t_{\mathrm{end}}$, when the scalar field becomes
equal to zero and then it becomes to oscillate. Because, as was
noted above, the duration of inflationary stage is greater in one
order or more than duration of transition stage, we can consider
the value of $ t_{\mathrm{end}}$ as estimation of duration of
inflationary stage. Numerical investigation of dependence of time
$t_{\mathrm{end}}$ on parameters $m$, $\veps$ and on initial
conditions for $\phi_0$, $S_{20}$, $S_{20}'$ leads us to the
following results. The value of $t_{\mathrm{end}}$ depends
essentially only on parameter $m$ and initial value of $\phi_0$,
and by given values of $\phi_0$ and $m$ it depends weakly on
admissible values of $\veps$ and initial values of $S_{20}$,
$S_{20}'$ (Figures~5--9).

\begin{figure}[hbt!]
\begin{minipage}{0.48\textwidth}\centering{
 \includegraphics[width=\linewidth]{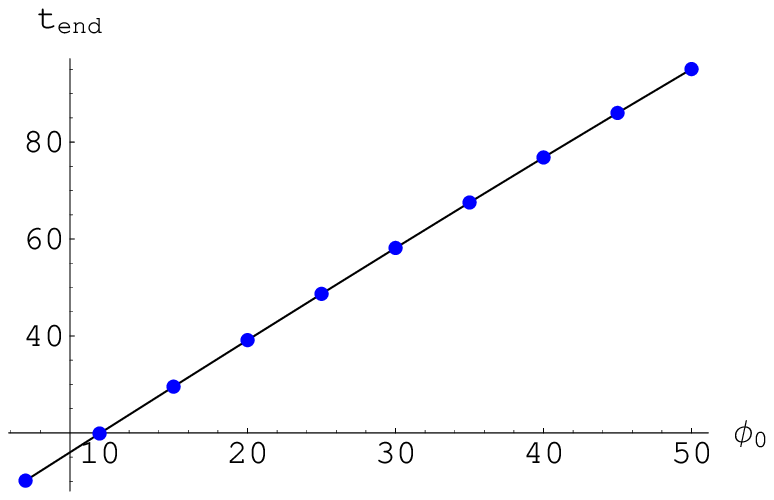}
}
 \caption[]{The dependence of $t_{\mathrm{end}}$ on $\phi_0$ ($\veps=10^{-5}$,
 $m=1{.}1$, $S_{20}=10^{-5}$, $S_{20}'=0$).}
\end{minipage}\, \hfill\,
\begin{minipage}{0.48\textwidth}\centering{
 \includegraphics[width=\linewidth]{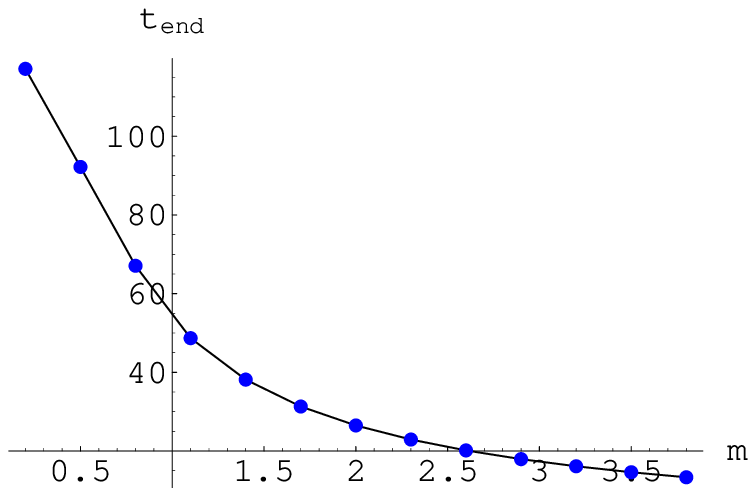}
}
 \caption[]{The dependence of $t_{\mathrm{end}}$ on $m$ ($\veps=10^{-5}$,
 $\phi_0=25$, $S_{20}=10^{-5}$, $S_{20}'=0$).}
\end{minipage}
\end{figure}

\begin{figure}[hbt!]
\begin{minipage}{0.48\textwidth}\centering{
 \includegraphics[width=\linewidth]{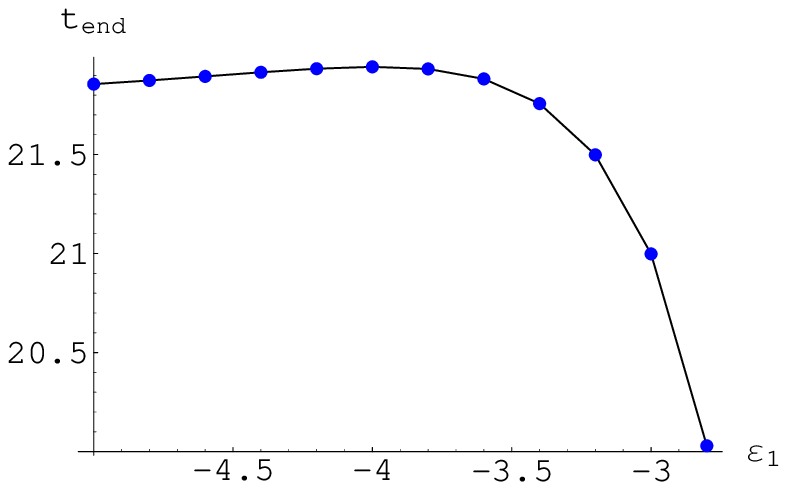}
}
 \caption[]{The dependence of $t_{\mathrm{end}}$ on $\veps$ ($\veps=10^{\veps_1}$,
 $\phi_0=10$, $m=1{.}0$, $S_{20}=10^{-5}$, $S_{20}'=0$).}
\end{minipage}\, \hfill\,
\begin{minipage}{0.48\textwidth}\centering{
 \includegraphics[width=\linewidth]{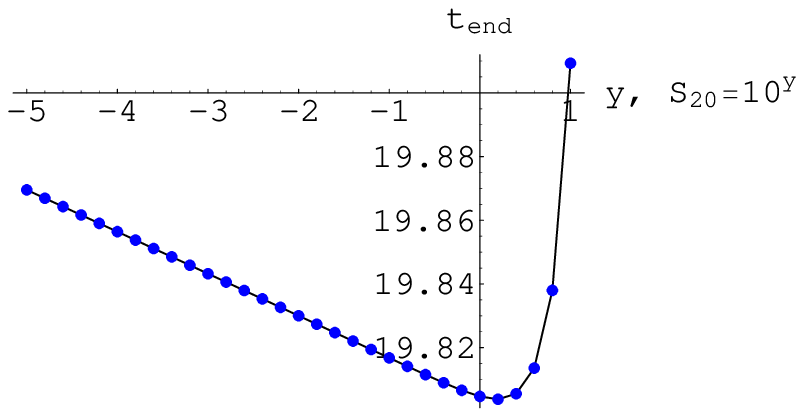}
}
 \caption[]{The dependence of $t_{\mathrm{end}}$ on $S_{20}$ ($\veps=10^{-5}$,
 $m=1.0$, $\phi_0=25$, $S_{20}'=10^{-4}$).}
\end{minipage}
\end{figure}

\begin{figure}[hbt!]
\begin{minipage}{0.48\textwidth}\centering{
 \includegraphics[width=\linewidth]{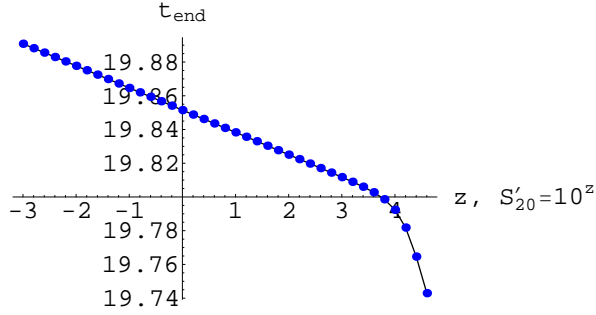}
}
 \caption[]{The dependence of $t_{\mathrm{end}}$ on $S_{20}'$ ($\veps=10^{-5}$,
 $\phi_0=10$, $m=1.0$, $S_{20}=10^{-5}$).}
\end{minipage}\, \hfill\,
\end{figure}

Similar to GR, the value of duration of inflationary stage depends
on initial value of $\phi_0$ in linear way (Fig.~5). From
numerical analysis of the value of $t_{\mathrm{end}}$ in
dependence on parameters and initial conditions follows:
\begin{enumerate}
\item similar to GR, there is a lower limit for initial value of
$|\phi_0|$ to have sufficient number of e-folds during the
inflationary stage ($|\phi_0|\approx 10$);
 \item for given value of $\veps$ admissible values for $|\phi_0|$
have upper limit;
 \item the parameter $\veps$ has upper limit depending on values $m$,
$\phi_0$ and $S_{20}'$ (in Fig.~7: $\veps\leq 10^{-3}$);
 \item there are upper limits for admissible values of $S_{20}$ and $S_{20}'$ at
given values of $m$, $\veps$ and $\phi_0$ (Fig.~8--9).
\end{enumerate}

Note that the presence of relativistic matter besides of scalar
field leads only to quantitative corrections and does not change
general conclusions given above. In particular, the presence of
ultrarelativistic matter at transition stage changes the amplitude
of oscillations $S_2'$-function. The influence of
ultrarelativistic matter at inflationary and postinflationary
stages is negligibly small because the energy density of
ultrarelativistic matter rapidly decreases during inflationary
stage.

The behaviour of $H$ and $\phi$ during the inflationary stage are
similar to that of GR. (Figures~10). There are small differences
for $H$ at the start and the end of inflationary stage. The graphs
in Fig.~10 and also in Fig.~11 for $S_2$-function are obtained for
$m=1.1$, $\veps=10^{-5}$, $\phi_0=10$, $S_{20}=0$,
$S_{20}'=10^{-4}$. During the inflationary stage the following
relations are satisfied with a rather high accuracy
$$
H\approx \sqrt{\frac{1}{2}m^2\phi^2 + \frac{1}{2}{\phi'}^2},
\qquad S_2\approx \sqrt{m^2\phi^2 - \frac{1}{2}{\phi'}^2}.
$$

\begin{figure}[hbt!]
\begin{minipage}{0.48\textwidth}\centering{
 \includegraphics[width=\linewidth]{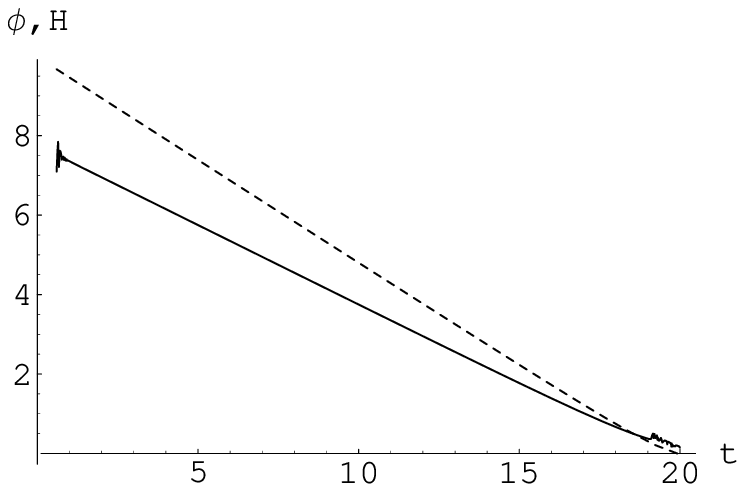}
} \caption[]{The behaviour of $H$ (firm line) and $\phi$ (dashed
line) during the inflationary stage.}
\end{minipage}\, \hfill\,
\begin{minipage}{0.48\textwidth}\centering{
 \includegraphics[width=\linewidth]{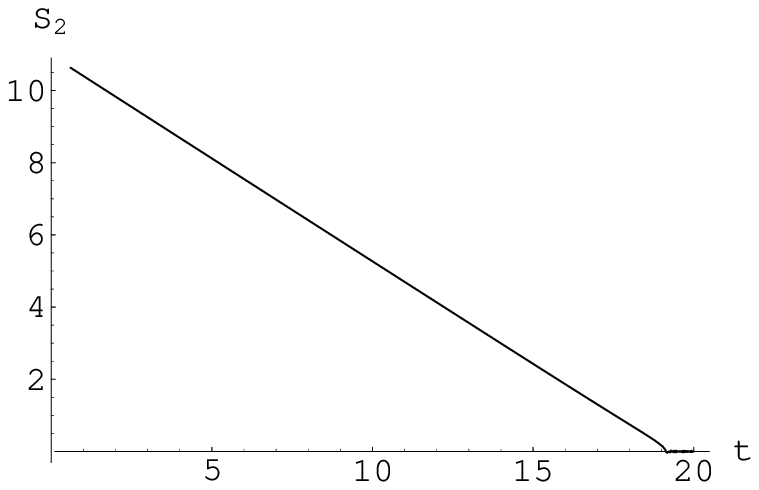}
} \caption[]{The behaviour of $S_2$ during the inflationary stage.}
\end{minipage}\\
\end{figure}

The behaviour of scalar field $\phi$, torsion function $S_2$ and
Hubble parameter $H$ during the postinflationary stage are
presented in Figures~12--14. The graphs in Figures~12, 13 and 14a
are obtained at the following values of parameters
$\veps=10^{-5}$, $m=1{.}1$ and initial conditions: $\phi_0=10$,
$S_{20}=0$ and $S_{20}'=10^{-4}$. Similar to GR during the
postinflationary stage the scalar field $\phi$ oscillates with
decreasing amplitude. The frequency of $\phi$-oscillations
increases by increasing of the parameter $m$. The oscillations of
the Hubble parameter have the character of beats (Fig.~14). Two
subsequent pulsation of beats are divided by domain of
oscillations with small amplitude. Similar to inflationary models
without torsion function $S_2$ investigated in [3], for
sufficiently large values of the parameter $m$ the oscillating
Hubble parameter changes its sign (in Figures 14a and 14b the
value of $m$ is equal to 1.1 and 0.4 respectively). In this case
the cosmological model vibrates during some time interval after
inflation. The amplitude of $H$-beats increases by increasing of
parameter $m$. By decreasing of $\veps$ the frequency of
oscillation inside one pulsation increases although the frequency
of beats oneself practically does not change. There is the
correlation between $\phi$-oscillations, $H$-beats and the
temporal behaviour of $S_2$-function during the postinflationary
stage. When instantaneous value of $\phi$ is small in absolute
value, then $S_2$-function oscillates near zero and $H$ decreases
being between two subsequent pulsations.  When absolute value of
$\phi$ is near to its maximum, the $H$-parameter oscillates inside
of pulsation and the absolute value of $S_2$-function increases.

\begin{figure}[htb!]
\begin{minipage}{0.48\textwidth}\centering{
 \includegraphics[width=\linewidth]{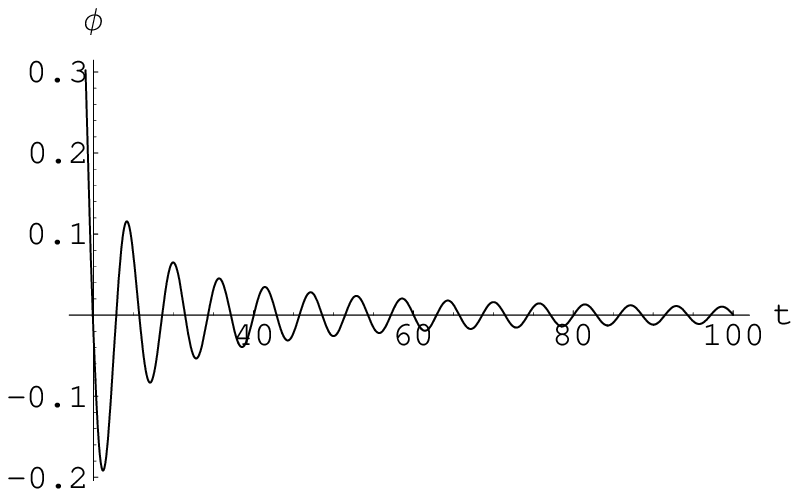}
} \caption[]{The behaviour of $\phi$ during the postinflationary
stage.}
\end{minipage}\, \hfill\,
\begin{minipage}{0.48\textwidth}\centering{
 \includegraphics[width=\linewidth]{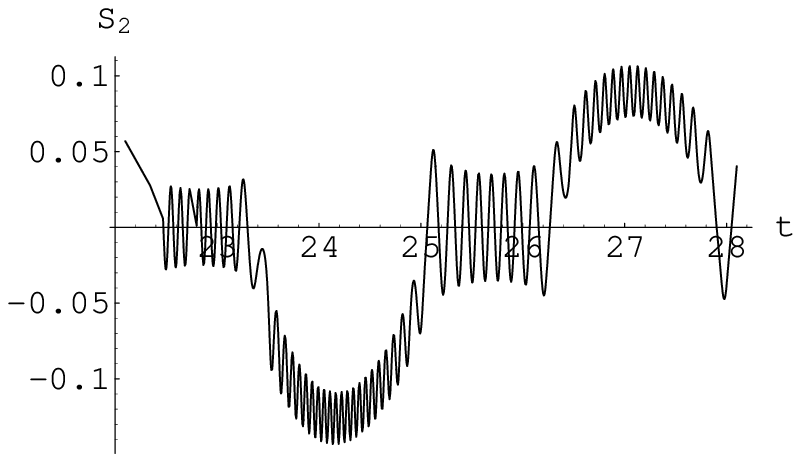}
} \caption[]{The behaviour of $S_2$ during the postinflationary
stage.}
\end{minipage}
\end{figure}

\begin{figure}[htb!]
\begin{minipage}{0.48\textwidth}\centering{
 \includegraphics[width=\linewidth]{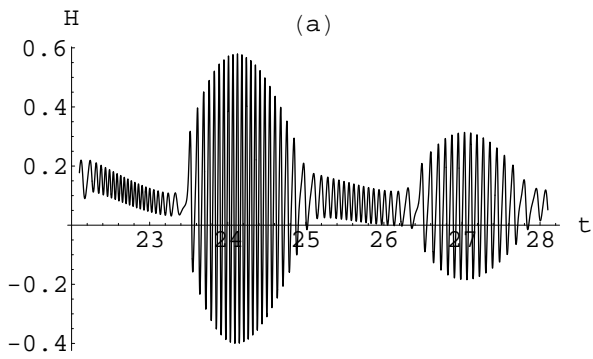}
} 
\end{minipage}\, \hfill\,
\begin{minipage}{0.48\textwidth}\centering{
 \includegraphics[width=\linewidth]{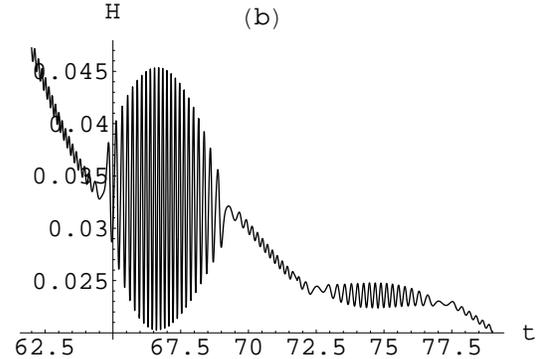}
} 
\end{minipage}
\caption[]{The behaviour of $H$ during the postinflationary
stage.}
\end{figure}

As result of this Section, we can conclude that physically
interesting solutions exist for some restrictions on parameters
$m$, $\veps$ and for sufficiently large domain of initial
conditions.

\section{Extremum surface for initial conditions}

It is interesting to investigate more particularly the structure
of extremum surface defined by (13), on which initial conditions
for regular inflationary cosmological solutions are given.
Supposing below $\rho=0$ and $k=0$, we will have extremum surface
in 4-dimensional space of variables ($\phi_0$, $\phi_0'$,
$S_{20}$, $S_{20}'$). By taking into account results concerning
the duration of time $ t_{\mathrm{end}}$ obtained in previous
Section, we will consider 3-dimensional subspace $P_3$ of
variables ($S_{20}$, $S_{20}'$, $\phi_0'$), then the relation (13)
determines 2-dimensional surface $P_2$ in $P_3$ depending
parametrically  on variable $\phi_0$ and also on parameters $m$
and $\veps$. The 2-dimensional surface $P_2$ has sufficiently
complicated structure and includes a closed cover  $P_{20}$ and
also some complicated surfaces surrounding the closed cover
$P_{20}$. A part of surface $P_2$ is presented in Figure~15.

As numerical analysis shows, we obtain regular inflationary
cosmological solutions by choosing initial conditions on closed
cover  $P_{20}$. The study of geometrical properties of surface
$P_2$ allows to understand some important properties of regular
inflationary cosmological solutions at transition stage from
compression to expansion. With this purpose we will consider
cross-sections of the surface $P_2$ with planes orthogonal to
coordinate axes $S_{20}$, $S_{20}'$ and $\phi_0'$. As example, in
Fig.~16 corresponding cross-section in the plane $S_{20}'=10^{-4}$
is presented ($m=1{.}1$, $\veps=10^{-5}$, $\phi_0=10$).

\begin{figure}[hbt!]
\begin{minipage}{0.48\textwidth}\centering{
 \includegraphics[width=\linewidth]{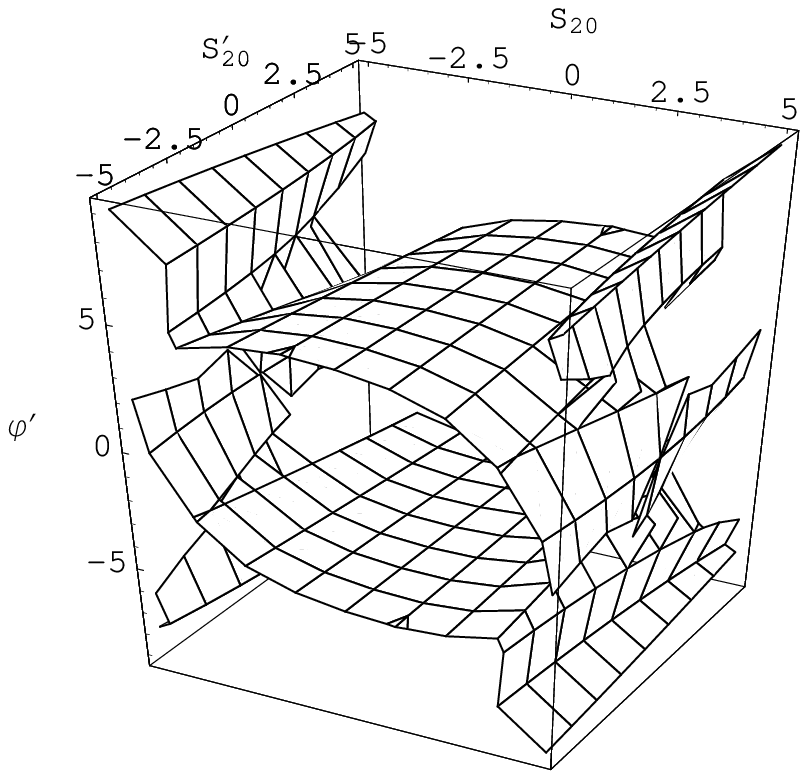}
} \caption[]{Part of extremum surface $P_{20}$.}
\end{minipage}\, \hfill\,
\begin{minipage}{0.48\textwidth}\centering{
 \includegraphics[width=\linewidth]{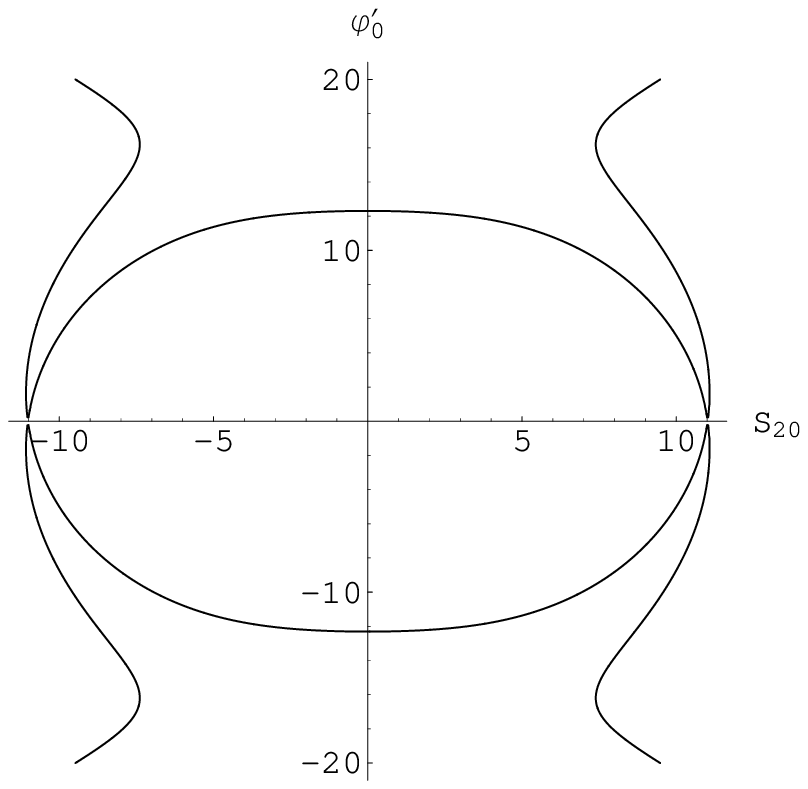}
} \caption[]{Cross-section of extremum surface $P_{20}$ by the
plane $S_{20}'=10^{-4}$.}
\end{minipage}\\
\end{figure}

By increasing of $m$ and the value of variable $\phi_0$ linear
sizes of the closed cover  $P_{20}$ increase. By decreasing of
$\veps$ linear sizes of $P_{20}$ increase only in directions of
axes $S_{20}$ and $S_{20}'$ and do not change practically in
direction of $\phi_0'$.

As it was discussed in previous Section, the Hubble parameter
oscillates at transition stage from compression to expansion. This
means that during this stage the quantities $H$, $S_2$, $S_2'$,
$\phi$, $\phi'$ are changing by intersecting the extremum surface
$H=0$ and by changing the sign of derivative $H'$. In order to
find domains on closed cover $P_{20}$, where the sign of $H'$ is
positive and negative, we have to consider the intersection of
$P_{20}$ with surface $H'=0$ defined by (12). The sign of $H'$ on
$P_{20}$ is different from different sides of intersection line.
The place of intersection of $P_{20}$ with surface $H'=0$ depends
on the value of $\phi_0$. So, in the case $\phi_0=10$ the domain
of intersection corresponds to small values of $S_{20}'$ and to
values of $S_{20}$ near to its maximum. By increasing of $\phi_0$
the domain of intersection moves in direction of decreasing values
of $S_{20}$ and increasing values of $S_{20}'$. The presence of
domains with positive and negative values of derivative $H'$ on
extremum surface $P_{20}$ allows to explain the oscillating
character of the Hubble parameter at transition stage, that is
connected essentially with behaviour of $S_2$-function (see
Fig.~3). After the bounce at $t=0$ the $S_2$-function quickly
increases and reaches its maximum. As a result, oscillating
function $H(t)$ reaches now extremum surface at value of $S_{20}$,
where the derivative $H_0'$ is negative. Together with changing of
$S_2$-function the process of oscillations of the Hubble parameter
is repeated at decreasing values of $\phi_0$ (see Fig.~2) and as
result at decreasing values of $S_{20\mathrm{max}}$. Oscillations
of $H$-function will continue even after this function does not
reach extremum surface. The amplitude of $H$-oscillations will
continue to decrease  until the inflationary stage will come.

\section{Conclusion}

As follows from our analysis, regular inflationary cosmological
solutions for HIM with two torsion functions built in the
framework of PGTG are realized at certain restrictions on
parameters of HIM and at large domain of initial conditions on
extremum surface $H=0$. The presence of pseudoscalar torsion
function $S_2(t)$ leads to essential changes of considered
solutions in comparison with inflationary cosmological solutions
for HIM without $S_2$-function analyzed in [3]. At first of all
these changes relate to properties of transition stage from
compression to expansion and post-inflationary stage and are
connected with oscillating behaviour of the Hubble parameter.
Differences of post-inflationary stages for cosmological solutions
in considered theory in comparison with general relativity theory
can lead to quantitative differences by transition to
radiation-dominated stage, in particular, to differences in
anisotropy of relic radiation. This means that the building of
perturbation theory for scalar fields in considered inflationary
cosmological HIM is of direct physical interest.


\begin{thebibliography}{99}
\bibitem{1} Minkevich A V 2006 {\it Gravitation\&Cosmology\/} \textbf{12} 11--21 ({\it Preprint\/} gr-qc/0506140)
\bibitem{2} Minkevich A V 2007 {\it Acta Physica Polonica B\/} \textbf{38} 61--72 ({\it Preprint\/} gr-qc/0512123)
\bibitem{3} Minkevich A V and Garkun A S 2006 {\it Class. Quantum Grav.\/} {\textbf{23}} 4237--47
 ({\it Preprint\/} gr-qc/0512130)
\bibitem{4} Minkevich A V, Garkun A S and Kudin V I 2007 {\it Class. Quantum Grav.\/} {\textbf{24}}
5835--47 ({\it Preprint\/} Arxiv:0706.1157)
\bibitem{5} Minkevich A V 2007 {\it Ann. Fond. Louis de Broglie \/} \textbf{32} 253--266 ({\it Preprint\/} Arxiv:0709.4337)
\end{thebibliography}
\end{document}